\documentclass[showpacs,amssymb,prd,superscriptaddress,twocolumn,aps,nofootinbib]{revtex4-1}
\usepackage{graphicx, epsfig, amssymb} 
\usepackage{amsmath, amsfonts}
\usepackage{bm} 
\usepackage[breaklinks]{hyperref}
\usepackage{color}

\def\be{\begin{equation}}
\def\ee{\end{equation}}
\def\beq{\begin{eqnarray}}
\def\eeq{\end{eqnarray}}

\begin{document}

\title{Constraining Primordial Black-Hole Bombs through Spectral Distortions of the Cosmic Microwave Background}

\author{Paolo Pani}\email{paolo.pani@ist.utl.pt}
\affiliation{CENTRA, Departamento de F\'{\i}sica, 
Instituto Superior T\'ecnico, Universidade T\'ecnica de Lisboa - UTL,
Avenida Rovisco Pais 1, 1049 Lisboa, Portugal.}
\affiliation{Institute for Theory and Computation, Harvard-Smithsonian
CfA, 60 Garden Street, Cambridge, MA 02138, USA}

\author{Abraham Loeb}\email{aloeb@cfa.harvard.edu}
\affiliation{Institute for Theory and Computation, Harvard-Smithsonian
CfA, 60 Garden Street, Cambridge, MA 02138, USA}

\begin{abstract} 
We consider the imprint of superradiant instabilities of nonevaporating primordial black holes (PBHs) on the spectrum of the cosmic microwave background (CMB). In the radiation dominated era, PBHs are surrounded by a roughly homogeneous cosmic plasma which endows photons with an effective mass through the plasma frequency. 
In this setting, spinning PBHs are unstable to a spontaneous spindown through the well-known``black-hole bomb'' mechanism. At linear level, the photon density is trapped by the effective photon mass and grows exponentially in time due to superradiance. 
As the plasma density declines due to cosmic expansion, the associated energy around PBHs is released and dissipated in the CMB. We evaluate the resulting spectral distortions of the CMB in the redshift range $10^3\lesssim z\lesssim 2\times10^6$. Using the existing COBE/FIRAS bounds on CMB spectral distortions, we derive upper limits on the fraction of dark matter that can be associated with spinning PBHs in the mass range $10^{-8}M_\odot\lesssim M \lesssim 0.2M_\odot$. For maximally-spinning PBHs, our limits are much tighter than those derived from microlensing or other methods. Future data from the proposed PIXIE mission could improve our limits by several orders of magnitude.
\end{abstract}


\maketitle


\section{Introduction}
Primordial black holes (PBHs) could have formed due to large density fluctuations in the early universe~\cite{Carr:2009jm,Carr:2005bd}. They acquire the mass contained within the particle horizon at the time they were formed, and PBHs formed in the first few seconds would span a wide range of masses, $10^{-5}{\rm g}<M<10^5 M_\odot$.
PBHs provide unique probes of early cosmology and high-energy physics; for example, PBH production could have been enhanced during phase transitions when the cosmic pressure suddenly declined~\cite{Jedamzik:1996mr} and it is very sensitive to non-Gaussianity~\cite{2013arXiv1307.4995Y}.  Since PBHs are collisionless and nonrelativistic they are natural dark matter (DM) candidates.

PBHs with mass $M\lesssim 10^{-18}M_\odot$ would have evaporated by the present time through their emission of Hawking radiation~\cite{Hawking:1974rv,Hawking:1974sw}. The lack of detected $\gamma$-rays from PBHs with $M\lesssim 10^{-18}M_\odot$ put very stringent constraints on their present density~\cite{Page:1976wx}. Hence, experimental and theoretical attempts to constrain PBHs focused on nonevaporating PBHs with mass $M>10^{-18}M_\odot$. Constraints on the DM fraction in PBHs were derived in the range $M>10^{-7}M_\odot$, based on dynamical~\cite{Quinn:2009zg,0004-637X-516-1-195}, microlensing~\cite{MACHO,EROS,GRB,QSO} and astrophysical~\cite{Ricotti:2007au,Mack:2006gz,Kesden:2011ij} effects (see Ref.~\cite{Carr:2009jm} for an overview). On the other hand, the interval $10^{-18}M_\odot\lesssim M\lesssim 10^{-7}M_\odot$ is still poorly constrained. In this mass range, light PBHs can satisfy Big Bang nucleosynthesis and Cosmic Microwave Background (CMB) limits, and make up the DM. Although more massive PBHs are ruled out as the sole DM constituents, they might still play an important role in cosmology; e.g. PBHs with $M\sim 10^3 M_\odot$ might seed the growth of supermassive BHs at redshift $z\gtrsim6$ due to accretion in the matter-dominated era~\cite{Mack:2006gz}.

In this paper we point out that if PBHs are formed with a nonvanishing spin then a novel mechanism can be used to derive very stringent theoretical bounds on their abundance in the mass range $10^{-8}M_\odot\lesssim M\lesssim 0.2 M_\odot$. All existing constraints on  PBHs ignored their spin; while this assumption is justified for evaporating PBHs (because spin is radiated faster than the mass~\cite{PhysRevD.14.3260}) we will show here that the spin of nonevaporating PBHs could affect dramatically their impact on the CMB. 

There is no fundamental reason to believe that PBHs are formed with a vanishing angular momentum. In fact, all mechanisms that were proposed for PBH formation in the early universe (e.g. bubble collisions, collapse of string loops, or density fluctuations during inflation ~\cite{Frampton:2010sw}) should naturally produce nonzero PBH spin. Thus, it is particularly important to understand how the inclusion of spin would modify current constraints on the DM fraction in PBHs.

An effect that comes into play when BHs possess nonzero angular momentum is superradiance~\cite{Teukolsky:1974yv}. A low-frequency bosonic wave scattered off a spinning BH is amplified when its frequency satisfies the superradiant condition $\omega<m\Omega_H$, where $m$ is the azimuthal number of the wave and $\Omega_H$ is the angular velocity of the BH horizon. In 1972 Press \& Teukolsky proposed that if a BH was surrounded by a reflecting surface, successive superradiant amplifications would trigger an instability, dubbed ``BH bomb''~\cite{Press:1972zz,Cardoso:2013zfa}. In this process, angular momentum is extracted from the BH until the superradiant condition is saturated. A natural way to provide successive reflections is when the BH interacts with a massive bosonic field~\cite{Damour:1976kh,Detweiler:1980uk,Cardoso:2004nk}, since in this case the mass may confine low-frequency perturbations in a region $\sim 1/\omega_p$. Massive standard-model particles (e.g. pions) might trigger superradiant instabilities of isolated PBHs with mass $M\lesssim10^{-18} M_\odot$~\cite{Detweiler:1980uk}. 

Interestingly, if the BH is not isolated but surrounded by a hot plasma, even photons acquire an effective mass given (in natural units $G=c=\hbar=1$) by the plasma frequency \cite{Sitenko:1967,Kulsrud:1991jt}
\begin{equation}
 \omega_p=\sqrt{{4\pi e^2 n}/{m_e}}\,,\label{plasma_freq}
\end{equation}
where $n$ is the electron density and $m_e$ and $e$ are the electron mass and charge, respectively. As shown below, the interaction with the plasma  spontaneously triggers a superradiant instability of spinning PBHs over a wide (and presently poorly constrained) range of masses. Because the mean density of the cosmic gas decreases as the universe expands, the instability of PBHs with different masses is effective at different redshifts.

\begin{figure}[thb]
\begin{center}
\begin{tabular}{cc}
 \epsfig{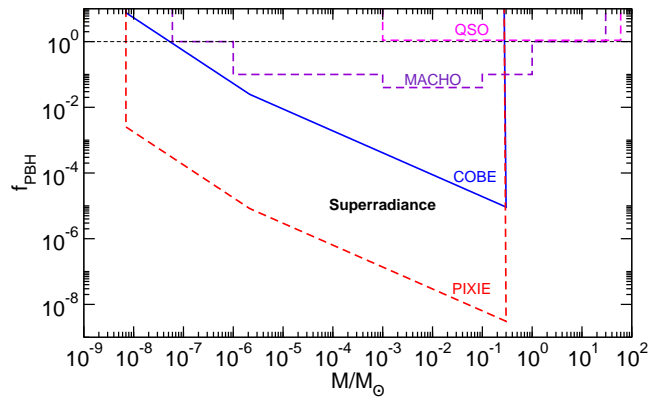}
\end{tabular}
\caption{Upper limits on the mass fraction of DM in PBHs with masses in the range $10^{-9}M_\odot<M<10^2 M_\odot$. 
The solid blue curve is the theoretical constraint derived in this paper using COBE/FIRAS data~\cite{Fixsen:1996nj}. The dashed red line is the expected limit from the proposed  PIXIE experiment~\cite{Kogut:2011xw}. Our limits are plotted for maximally-spinning PBHs with $\langle \tilde{a}\rangle=1$, and scale roughly as $1/\langle \tilde{a}\rangle$ (see text for details). The limits from other methods are adopted from Ref.~\cite{Carr:2009jm}.
  \label{fig:constraints} }
\end{center}
\end{figure}

As a result of the instability, PBHs transfer part of their angular momentum and mass to the electromagnetic (EM) field, whose energy density grows exponentially in their  vicinity (see Ref.~\cite{Arvanitaki:2010sy} for a similar discussion in the case of axions around astrophysical BHs). As the gas density declines during cosmic expansion, the associated energy  is dissipated into the CMB, potentially leading to spectral distortions from a perfect blackbody spectrum \cite{Sunyaev:2013aoa}. Using COBE/FIRAS data~\cite{Fixsen:1996nj}, we have estimated the upper bounds on the DM fraction in spinning PBHs. Our results are summarized in Fig.~\ref{fig:constraints} and discussed in detail through the rest of the paper.

\section{Plasma-triggered superradiant instabilities}
We consider a spinning BH surrounded by a plasma. If the total mass of the surrounding matter is sufficiently small, its gravitational backreaction is negligible and the background spacetime is uniquely described by the Kerr metric. The latter is defined by only two parameters, the mass $M$ and the dimensionless spin parameter $\tilde{a}\equiv J/M^2$, where $J$ is the BH angular momentum.
%
Photons interacting with the plasma acquire an effective mass given by Eq.~\eqref{plasma_freq}~\cite{Sitenko:1967}. As a consequence of the modified dispersion relation, Maxwell equations within the plasma in flat spacetime read
\begin{equation}
\nabla_\sigma F^{\sigma\nu}=\omega_p^2 A^\nu\,,\label{proca}
\end{equation}
where $F_{\mu\nu}=\partial_\mu A_\nu-\partial_\nu A_\mu$ and $A_\mu$
is the vector potential. The equation above is also valid in curved spacetime as long as the background is slowly varying compared to $\omega_p^{-1}$ and the density gradient is small compared to the gravitational field~\cite{Kulsrud:1991jt}.

Equation~\eqref{proca} has been studied in detail on a Kerr background for $\omega_p=$const, when it coincides with the  well-known Proca equation governing the dynamics of a massive spin-1 field in vacuum. The same equation governs the dynamics of standard (massless) photons which acquire an effective mass due to their interaction with a homogeneous plasma.

For $\omega_p\neq0$ Eq.~\eqref{proca} does not admit separation of variables in the Kerr background and one has to resort to approximate schemes in the frequency domain~\cite{Pani:2012bp,Pani:2012vp} or to a time evolution~\cite{Witek:2012tr}. 
As shown in Refs.~\cite{Pani:2012bp,Pani:2012vp},
at linear level the system develops a superradiant instability, similar to that occurring for a massive scalar field around a Kerr BH~\cite{Damour:1976kh,Detweiler:1980uk,Cardoso:2004nk,Dolan:2007mj,Arvanitaki:2010sy}. The instability is regulated by the dimensionless parameter $M\omega_p$ and it is maximum when $M\omega_p\sim0.4$ and for nearly-extremal BHs~\cite{Witek:2012tr}.

It is particularly convenient to estimate the instability timescale in the frequency domain. Fourier decomposing the fields as $A_\mu(t,\vec{x})=\int d\omega e^{-i\omega t}\tilde{A}_\mu(\omega,\vec{x})$, where $\omega=\omega_R+i\omega_I$ and using a slow-rotation framework, we can estimate the frequency $\omega_R$ of the unstable modes and the instability timescale $\tau_{\rm SR}=\omega_I^{-1}$ in the small $M\omega_p$ limit as follows~\cite{Pani:2012vp}
\begin{eqnarray}
 \omega_R^2&\sim&\omega_p^2\left[1-\left(\frac{M\omega_p}{\ell+N+S 
+1}\right)^2\right]+{\cal O}\left(\omega_p^4\right)\,,\label{fit_wR}\\
M\tau_{\rm SR}^{-1}&\sim& \gamma_{S\ell}\left(\tilde{a}m-2 r_+\omega_p \right) (M\omega_p)^{4\ell+5+2S}\,, \label{fit}
\end{eqnarray}
where $r_+$ is the horizon radius, $\ell$ is the harmonic index of the corresponding mode, $N$ is an integer, $S=\pm1,0$ is the mode polarization and $\gamma_{S\ell}$ is a numerical coefficient. Although the above results have been derived to second order in $\tilde{a}$, numerical simulations in the near-extremal $(\tilde{a}=0.99)$ regime agree with an extrapolation of the above formulae to within a factor 2~\cite{Witek:2012tr}.

The time-averaged angular momentum flux $\dot L_H$ and energy flux $\dot E_H$ across the BH horizon satisfy~\cite{Wald:1984rg,Arvanitaki:2010sy}

\begin{equation}
 \dot L_H= \frac{m}{\omega_R}\dot E_H\propto m(\omega_R-m\Omega_H)|\Psi(r,\vartheta)|^2\,, \label{dotL}
\end{equation}
where $\Omega_H={\tilde{a}}/({2r_+})$ and $\Psi(r,\vartheta)$ are the angular and radial components of the field amplitude in Fourier space, whose time dependence is exponential, $\propto e^{t/{\tau_{\rm SR}}}$, in the superradiant regime.
Equation~\eqref{dotL} implies that if the superradiant condition is met, $\omega_R<m\Omega_H$, the angular momentum and energy fluxes across the horizon are negative, i.e. the instability extracts energy from the BH and transfers it to the EM field~\cite{Arvanitaki:2010sy}.

\section{Primordial Black-Hole Bombs}
PBHs formed in the early universe are surrounded by a mean cosmic electron density,
\begin{equation}
 n=n_0\left(1+z\right)^3\approx 220{\rm cm}^{-3}\left(\frac{1+z}{10^3}\right)^3\,,\label{cosmicdensity}
\end{equation}
as a function of redshift $z$, which translates to a time-dependent plasma frequency through Eq.~\eqref{plasma_freq}. If the cosmological evolution occurs on a much longer timescale than the BH evolution, we can  adopt an adiabatic approximation and treat $n$ as constant during the energy extraction phase at a given $z$. In the following, we assume that the plasma density near the BH is constant, homogeneous and given by Eq.~\eqref{cosmicdensity} (see \S~\ref{sec:discussion} for a discussion). Under these assumptions, we can directly apply the results of Refs.~\cite{Pani:2012bp,Pani:2012vp,Witek:2012tr} for a Proca field with mass $\omega_p$ around an isolated Kerr BH.

\begin{figure}[thb]
\begin{center}
\begin{tabular}{cc}
 \epsfig{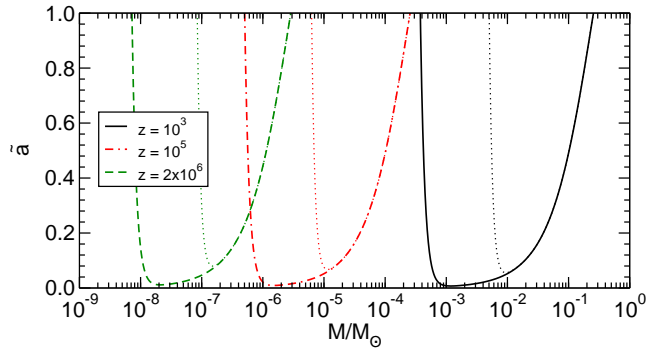}
\end{tabular}
\caption{Contour plots in the BH Regge plane~\cite{Arvanitaki:2010sy,Pani:2012bp} corresponding to an instability timescale shorter than the age of the universe at a given redshift, assuming a mean cosmic gas density as in Eq.~\eqref{cosmicdensity}. Thick and thin curves correspond to polar and axial modes, respectively~\cite{Pani:2012bp}. In both families, the rightmost part of each curve is described by $\tilde{a}\sim 4 M \omega_p$. Roughly speaking, PBHs in the mass range $7\times 10^{-9}M_\odot<M<0.2 M_\odot$ go through a cosmic era (at some $10^3<z<2\times10^6$) when the superradiant instability is effective.
  \label{fig:ReggePlane_PBHs} }
\end{center}
\end{figure}

In order for the superradiance instability to be effective at a given redshift $z$, the instability timescale must be much shorter than the cosmological evolution timescale.
By comparing the timescale $\tau_{\rm SR}$ in Eq.~\eqref{fit} (with $l=m=1$) with the age of the Universe  $\tau_{\rm age}$ as a function of redshift, we show in Fig.~\ref{fig:ReggePlane_PBHs} the Regge plane~\cite{Arvanitaki:2010sy,Pani:2012bp} for PBHs with mass in the range $10^{-9} M_\odot<M< M_{\odot}$ for three representative redshift values. For the updated set of cosmological parameters~\cite{Ade:2013zuv}, $\tau_{\rm age}= 4.3\times 10^5$yr, $74.9$yr and $0.19$yr at $z=10^3$, $z=10^5$ and $z=2\times 10^6$, respectively. At any plotted $z$, PBHs located above the corresponding curve are unstable due to superradiant instability with $\tau_{\rm SR}<\tau_{\rm age}$.
Figure~\ref{fig:ReggePlane_PBHs} shows the fundamental polar modes (which have the shortest instability timescale) and the axial modes, for which an analytical expression for the timescale in the $M\omega_p\ll1$ limit is available~\cite{Pani:2012vp}. In both families, the rightmost part of each curve is perfectly described by $\tilde{a}\sim 4 M \omega_p$.
This threshold marks the portion of the Regge plane where superradiant instability starts becoming effective. Using Eqs.~\eqref{plasma_freq} and~\eqref{cosmicdensity}, we can translate this condition into an upper bound on the mass:
\begin{equation}
 \frac{M}{M_\odot}<0.19\tilde{a}\left(\frac{1+z}{10^3}\right)^{-3/2}\,. \label{threshold}
\end{equation}
Note that this result is very general and does not depend on the details of the superradiant spectrum, but only on the superradiant condition, $\omega_R<m\Omega_H$ and on the fact that the modes have a hydrogenic spectrum, $\omega_R\sim\omega_p$~\cite{Pani:2012vp}.
In other words, a PBH with mass $M$ and spin $\tilde{a}$ satisfying the relation above will pass through an epoch at redshift $z$ when the mean gas density is such that the superradiant instability is effective. 
%

\noindent{\bf{\em Energy and angular momentum extraction.}}
As a result of the superradiant instability, a spinning PBH could lose most of its spin energy over a short time scale. We note that the threshold curves shown in Fig.~\ref{fig:ReggePlane_PBHs} are very steep functions of the dimensionless angular momentum $\tilde{a}$ and extend almost down to $\tilde{a}=0$, i.e. the superradiant instability is active even for very small (but nonvanishing) BH spin. 
Thus, before superradiance stops being effective, a single PBH with initial dimensionless angular momentum $\tilde{a}$ will essentially lose all its angular momentum, i.e. $\Delta(\tilde{a})\approx\tilde{a}.$
Using this result and Eq.~\eqref{dotL} with $m=1$, we obtain
\begin{equation}
 \frac{\Delta M}{M}\approx\frac{\tilde{a} M\omega_R}{1-2\tilde{a} M\omega_R}\approx 10^{-3} \left(\frac{1+z}{10^3}\right)^{3/2}\left(\frac{\tilde{a}M}{10^{-3}M_\odot}\right) \,,\label{deltaM}
\end{equation}
where in the last step we assumed $(M/M_\odot)\ll 2\times 10^{5}(1+z)^{-3/2}$, a condition valid throughout the entire range considered here. According to this estimate, in the linear approximation the efficiency of the energy extraction at $z\sim10^5$ for $M\sim 10^{-4} M_\odot$ is roughly $\tilde{a} \times20\%$.

\noindent{\bf{\em Dissipation.}}
The extracted energy is transferred to the energy density of the EM field, which grows exponentially near the BH on a timescale $\tau_{\rm SR}$~\cite{Arvanitaki:2010sy}. As long as the plasma frequency is comparable to or higher than the mode wavelength, $\omega_p\gtrsim \omega_R$, the photon energy is trapped by the effective potential well of the plasma. As the Universe expands, the plasma density decreases according to Eq.~\eqref{cosmicdensity}. Once $\omega_p$ becomes smaller than $\omega_R$, the EM energy accumulated through superradiance is released.
This argument assumes that $\omega_R$ is not redshifted during the cosmic expansion. In addition, energy can be also released by photon outflow due to the exponentially-growing radiation pressure near the BH.

The released energy can be dissipated through various channels. The most efficient process involves Coulomb collisions, which yields a dissipation timescale~\cite{Kuzelev:2009}
\begin{equation}
 \tau_C=\frac{2}{\nu_{\rm ei}}\left(1+\frac{k^2}{\omega_p^2}\right)\,,
\end{equation}
for a transverse EM wave, where $\vec{k}$ is the wave vector, $\nu_{\rm ei}=2.91\times 10^{-6}{(n_e/{\rm cm^{-3}})}{\rm ln}\Lambda\left(T_e/{\rm eV}\right)^{-3/2} {\rm s}^{-1}$
is the electron-ion collision frequency, ${\rm ln}\Lambda$ is the Coulomb Logarithm and $T_e=2.726 (1+z)$K is the plasma temperature. For $k^2\ll\omega_p^2$ (which is valid in the adiabatic regime), Eq.~\eqref{cosmicdensity} yields the dissipation timescale:
\begin{equation}
 \tau_C\approx 15.5\left(\frac{1+z}{10^3}\right)^{-3/2}{\rm s}\,, \label{tauD}
\end{equation}
with $\log\Lambda\approx23$. This timescale is much shorter than the cosmic expansion time at the relevant redshifts.

\noindent{\bf{\em Spectral distortions and constraints.}}
An injection of an energy density $\Delta U$ into the CMB at redshifts $z\lesssim 2\times 10^6$ cannot be fully thermalized and results in observable distortions of the CMB spectrum from a perfect blackbody shape (see, e.g. Ref.~\cite{1991ApJ...379....1B}).

For simplicity, we focus on $\mu$ and $y$-type distortions. The former is produced by an energy injection $\Delta U$ at $10^5<z<2\times 10^6$ which introduces a chemical potential $\mu\sim 1.4 \Delta U/U$ to the Bose-Einstein spectrum, where $U$ is the unperturbed energy density of the CMB. At $z<10^5$, Compton scattering is unable to maintain a Bose-Einstein spectrum and $\mu$-distortions cannot be created. In the interval $10^3<z<10^5$ the energy injection produces $y$-distortions with amplitude $y=\Delta U/(4U)$. Other types of spectral distortions, such as intermediate $i$-type distortions occurring at $1.4\times 10^4<z <10^5$ can also be considered but computing their amplitude requires a case-by-case analysis (see e.g. Ref.~\cite{Khatri:2013dha}). After recombination ($z<10^3$) the Universe is optically thick to Compton scattering and essentially all energy is absorbed by the cosmic gas. Therefore, we are interested in the interval $10^3\lesssim z \lesssim 2\times 10^6$. This is different from the case of spectral distortions due to accretion onto PBHs~\cite{Ricotti:2007au}. In the latter case, since accretion is negligible in the radiation-dominated era, only the interval $3600\gtrsim z\gtrsim 1000$ is relevant and only $y$-type distortions can be produced.

The energy~\eqref{deltaM} extracted from each spinning PBH is injected into the CMB over the timescale much shorter than the cosmological evolution timescale at $10^3<z <2\times 10^6$. Using Eqs.~\eqref{plasma_freq}, \eqref{cosmicdensity}, \eqref{deltaM} and $U= \sigma T_e^4$, we obtain:
\begin{equation}
\frac{\Delta U}{U}=\langle \tilde{a}\rangle f_{\rm PBH} M \frac{\rho_{\rm crit}^0\Omega_{\rm DM}}{\sigma T_0^4}\sqrt{\frac{4\pi e^2 n_0}{m_e}}\left(1+z\right)^{1/2}\,,\label{dUoU}
\end{equation}
%
where $\sigma$ is the Stefan-Boltzmann constant, $f_{\rm PBH}$ is the present DM fraction in PBHs, $\Omega_{\rm DM}$ is the present-day DM density in units of the critical density $\rho_{\rm crit}^0=3H_0^2/(8\pi)$, $H_0=100h\,{\rm km}\,{\rm s}^{-1} {\rm Mpc}^{-1}$ is the Hubble constant, and 
\begin{equation}
 \langle \tilde{a}\rangle=\frac{\int_0^1 d\tilde a\, \tilde a \frac{dN}{d\tilde a}}{\int_0^1 d\tilde a\, \frac{dN}{d\tilde a}}\,, \label{average}
\end{equation}
is the average spin parameter weighted by the initial spin distribution of PBHs, $dN/d\tilde a$. 

Given a bound on $\Delta U/U$, Eq.~\eqref{dUoU} translates into a constraint on the combination $f_{\rm PBH}\langle \tilde{a}\rangle$. The $95\%$ confidence level limits from COBE/FIRAS  on the $\mu$ and $y$ parameters are $\mu<9\times 10^{-5}$ and $y<1.5\times 10^{-5}$, respectively~\cite{Fixsen:1996nj}. It follows that $\Delta U/U<6\times 10^{-5}$ at $95\%$ confidence level in the entire interval $10^3\lesssim z \lesssim 2\times 10^6$. Figure~\eqref{fig:constraints} shows the upper limits on the present DM fraction in PBHs for $\Omega_{\rm DM}\approx0.12/h^2$, $h=0.67$~\cite{Ade:2013zuv}, $\langle \tilde{a}\rangle=1$ and $z=\min(2\times 10^6,z_{\rm lim})$, with $z_{\rm lim}$ being the value that saturates Eq.~\eqref{threshold} for a given $M$ and $\tilde{a}$. The latter condition enforces the fact that any energy injection at $z\gtrsim2\times 10^6$ is fully thermalized. 

In addition to the term $\langle \tilde{a}\rangle$ in Eq.~\eqref{dUoU}, Eq.~\eqref{threshold} introduces a further dependence on $\tilde a$. It is easy to show that our bounds are only mildly affected by the spin dependence in Eq.~\eqref{threshold} (at least when $0.1<\tilde{a}<1$), so that the bounds on $f_{\rm PBH}$ roughly scale as $1/\langle \tilde{a}\rangle$.

The mass range in Fig.~\eqref{fig:constraints} is truncated at $M\sim0.2 M_\odot$ and $M\sim7\times 10^{-9} M_\odot$. The upper cutoff corresponds to the mass that saturates Eq.~\eqref{threshold} at $z=10^3$ and $\tilde{a}=1$. If $M\gtrsim0.2 M_\odot$, the energy is released after recombination when the Universe is optically thin to Thomson scattering and the coupling to the CMB is weak. The lower cutoff corresponds to the mass at which $\tau_{\rm SR}\approx \tau_{\rm age}$ at $z=2\times 10^6$. If $M\lesssim 7\times 10^{-9} M_\odot$, the instability timescale is long compared to the age of the Universe at $z=2\times 10^6$. 
\section{Discussion and Conclusions}\label{sec:discussion}

Figure~\eqref{fig:constraints} shows that superradiant instabilities of PBHs place very stringent bounds on the DM mass fraction $f_{\rm PBH}$ associated with maximally-spinning PBHs, over a mass range which was less severely constrained by other techniques such as microlensing. The solid blue line is the bound on $f_{\rm PBH}$ derived from the COBE/FIRAS data. The proposed PIXIE experiment~\cite{Kogut:2011xw} can drastically improve the limits on spectral distortions of the CMB. As recently shown~\cite{Chluba:2013dza,Kogut:2011xw}, such experiment has the potential to constrain $\mu$ and $y$ at the level of $1.4\times 10^{-8}\alpha$ and $1.2\times 10^{-9}\alpha$, respectively, where $\alpha~\sim{\cal O}(1)$ is a parameter that depends on the actual channel sensitivity of the experiment. Adopting $\alpha\sim1$ and using the most conservative of these constraints, a projected bound from a PIXIE-like experiment is $\Delta U/U<2\times 10^{-8}$ at the $95\%$ confidence level. This translates to the constraint on $f_{\rm PBH}$ shown by a red dashed curve in Fig.~\eqref{fig:constraints}. A PIXIE-like experiment can potentially improve the limit coming from COBE data by several orders of magnitude. In deriving the limits shown in Fig.~\eqref{fig:constraints} we have assumed that any energy injection  at $z>2\times 10^6$ is fully thermalized. 

The main limitation of the bounds derived through superradiant instabilities is their dependence on the BH spin. All other upper limits shown in Fig.~\eqref{fig:constraints} (cf. Ref.~\cite{Carr:2009jm}) are derived for nonspinning PBHs, so they actually constrain the quantity $f_{\rm PBH}$ alone, and not the combination $f_{\rm PBH}\langle \tilde a \rangle$. Nevertheless, our bounds can be considerably more stringent than the MACHO/EROS microlensing limits~\cite{MACHO,EROS} in the interval $10^{-6}M_\odot$ to $0.1 M_\odot$ and remain competitive even when $\langle \tilde{a}\rangle\sim0.01$. As an example, for a normal distribution centered at $\tilde{a}=0.3$ with width $\sigma= 0.1$, the average spin is $\langle \tilde{a}\rangle\sim0.3$. Production of nonspinning PBHs in the early Universe would  require fine-tuned initial configurations which are perfectly spherically-symmetric.

Although not shown in Fig.~\eqref{fig:constraints}, future experiments may be able to place competitive upper limits in a regime which partially overlaps with our theoretical bounds. The Kepler mission can constrain the range $5\times 10^{-10}M_\odot$ to $10^{-4}M_\odot$ through a search for microlensing signals~\cite{2011PhRvL.107w1101G,2013ApJ...767..145C}, and future pulsar-timing-array facilities like the Square Kilometer Array might be able to set a limit $f_{\rm PBH}>0.01$ in the range $5\times 10^{-12}M_\odot$ to $5\times 10^{-6}M_\odot$~\cite{Kashiyama:2012qz}. Constraints on PBHs in globular clusters~\cite{Capela:2013yf} might also be competitive, but are weakened by the lack of evidence for DM in these systems.

In our estimate we have neglected accretion onto PBHs. While accretion is negligible in the radiation-dominated era ($z\gtrsim 3.6\times 10^3$)~\cite{Ricotti:2007au}, it has to be considered in the interval $3.6\times 10^3\gtrsim z \gtrsim 10^3$. Nonetheless, as shown in Fig.~\eqref{fig:ReggePlane_PBHs}, the timescale of superradiant instabilities at $z\sim10^3$ is still much shorter than the accretion time ($\sim 4\times 10^7$yr). 
Even if accretion can be safely neglected at $z>10^3$, local inhomogeneities‎ of the plasma density can in principle affect our limits. Strictly speaking, Eqs.~\eqref{fit_wR} and \eqref{fit} are valid only when $\omega_p=$const. In a more realistic situation the plasma will have an inhomogeneous distribution due to the local gravitational field near the BH. In particular, the density is peaked at a few Schwarzschild radii whereas is negligible near the horizon. 

However, local density enhancements are weak during the radiation dominated era. The BH mass equals the background mass at the radius $R=(3\pi M\rho_{\rm crit}/4)^{1/3}$, where $\rho_{\rm crit}=3H^2/(8\pi)$ is the critical density at the corresponding redshift and $H=H(z)$ is the Hubble parameter. The matter outside $R$ feels a linear perturbation due to the BH. Since linear perturbations grow only logarithmically with time during the radiation dominated era, we may ignore the density enhancement outside $R$. Assuming Bondi accretion in the interior of $R$, the radius of influence of the BH is $R_{\rm acc}\sim M/c_s^2$ where $c_s$ is the speed of sound. In the radiation-dominated era, $c_s\sim1/\sqrt{3}$, so that $R_{\rm acc}$ is of order the Schwarzschild radius. This implies that the background cosmic density will only be enhanced by a factor of order unity near the BH.

Although detailed matter-distribution models are necessary for a quantitative assessment, using the methods developed in Ref.~\cite{Pani:2012vp} we have checked that the frequency and the timescale of the instability are insensitive to local inhomogeneities near the horizon, at least in the $M\omega_p\ll1$ limit which is relevant for our analysis. In fact, a detailed numerical study shows that Eq.~\eqref{fit} provides an upper limit for the instability timescale, and more realistic radial distributions would trigger even stronger instabilities.
We note that if the local plasma density near the BH is larger than the mean cosmic value by a factor $\gamma$, the upper limit on $f_{\rm PBH}\langle \tilde a \rangle$ would scale roughly as $\gamma^{-1}$. Our estimate is conservative as we assume $\gamma=1$.

While the assumption that the plasma frequency is constant in time is well justified when $\tau_{\rm SR}\ll\tau_{\rm age}$,  it would be interesting to extend recent computations in the frequency~\cite{Pani:2013pma} and time domains~\cite{Witek:2012tr} to obtain more precise estimates for the instability timescale triggered by a weakly inhomogeneous and time-dependent plasma distribution around a PBH.
Finally, fully-numerical evolutions~\cite{Witek:2012tr,Dolan:2012yt} may give us a better understanding of the nonlinear development and of the end state of the instability for generic spin.

The novel upper limits shown in Fig.~\eqref{fig:constraints} imply that
BH spin effects, combined with a background plasma, can dramatically affect the dynamics of nonevaporating PBHs and should be carefully considered in future studies.

\begin{acknowledgments}
We thank Vitor Cardoso and Helvi Witek for comments on the manuscript and useful conversations.
P.P. acknowledges financial support provided by the European Community through
the Intra-European Marie Curie contract aStronGR-2011-298297, by the NRHEP 295189 FP7-PEOPLE-2011-IRSES Grant, and by FCT-Portugal through the project CERN/FP/123593/2011. A.L. was supported in part by NSF grant AST-0907890 and NASA grants NNX08AL43G and NNA09DB30A.
\end{acknowledgments}
\bibliography{PBHs_superradiance}

\begin{thebibliography}{46}%
\makeatletter
\providecommand \@ifxundefined [1]{%
 \@ifx{#1\undefined}
}%
\providecommand \@ifnum [1]{%
 \ifnum #1\expandafter \@firstoftwo
 \else \expandafter \@secondoftwo
 \fi
}%
\providecommand \@ifx [1]{%
 \ifx #1\expandafter \@firstoftwo
 \else \expandafter \@secondoftwo
 \fi
}%
\providecommand \natexlab [1]{#1}%
\providecommand \enquote  [1]{``#1''}%
\providecommand \bibnamefont  [1]{#1}%
\providecommand \bibfnamefont [1]{#1}%
\providecommand \citenamefont [1]{#1}%
\providecommand \href@noop [0]{\@secondoftwo}%
\providecommand \href [0]{\begingroup \@sanitize@url \@href}%
\providecommand \@href[1]{\@@startlink{#1}\@@href}%
\providecommand \@@href[1]{\endgroup#1\@@endlink}%
\providecommand \@sanitize@url [0]{\catcode `\\12\catcode `\$12\catcode
  `\&12\catcode `\#12\catcode `\^12\catcode `\_12\catcode `\%12\relax}%
\providecommand \@@startlink[1]{}%
\providecommand \@@endlink[0]{}%
\providecommand \url  [0]{\begingroup\@sanitize@url \@url }%
\providecommand \@url [1]{\endgroup\@href {#1}{\urlprefix }}%
\providecommand \urlprefix  [0]{URL }%
\providecommand \Eprint [0]{\href }%
\providecommand \doibase [0]{http://dx.doi.org/}%
\providecommand \selectlanguage [0]{\@gobble}%
\providecommand \bibinfo  [0]{\@secondoftwo}%
\providecommand \bibfield  [0]{\@secondoftwo}%
\providecommand \translation [1]{[#1]}%
\providecommand \BibitemOpen [0]{}%
\providecommand \bibitemStop [0]{}%
\providecommand \bibitemNoStop [0]{.\EOS\space}%
\providecommand \EOS [0]{\spacefactor3000\relax}%
\providecommand \BibitemShut  [1]{\csname bibitem#1\endcsname}%
\let\auto@bib@innerbib\@empty
\bibitem [{\citenamefont {Carr}\ \emph {et~al.}(2010)\citenamefont {Carr},
  \citenamefont {Kohri}, \citenamefont {Sendouda},\ and\ \citenamefont
  {Yokoyama}}]{Carr:2009jm}%
  \BibitemOpen
  \bibfield  {author} {\bibinfo {author} {\bibfnamefont {B.}~\bibnamefont
  {Carr}}, \bibinfo {author} {\bibfnamefont {K.}~\bibnamefont {Kohri}},
  \bibinfo {author} {\bibfnamefont {Y.}~\bibnamefont {Sendouda}}, \ and\
  \bibinfo {author} {\bibfnamefont {J.}~\bibnamefont {Yokoyama}},\ }\href
  {\doibase 10.1103/PhysRevD.81.104019} {\bibfield  {journal} {\bibinfo
  {journal} {Phys.Rev.}\ }\textbf {\bibinfo {volume} {D81}},\ \bibinfo {pages}
  {104019} (\bibinfo {year} {2010})},\ \Eprint {http://arxiv.org/abs/0912.5297}
  {arXiv:0912.5297 [astro-ph.CO]} \BibitemShut {NoStop}%
\bibitem [{\citenamefont {Carr}(2004)}]{Carr:2005bd}%
  \BibitemOpen
  \bibfield  {author} {\bibinfo {author} {\bibfnamefont {B.~J.}\ \bibnamefont
  {Carr}},\ }\href@noop {} {\bibfield  {journal} {\bibinfo  {journal} {eConf}\
  }\textbf {\bibinfo {volume} {C041213}},\ \bibinfo {pages} {0204} (\bibinfo
  {year} {2004})},\ \Eprint {http://arxiv.org/abs/astro-ph/0504034}
  {arXiv:astro-ph/0504034 [astro-ph]} \BibitemShut {NoStop}%
\bibitem [{\citenamefont {Jedamzik}(1997)}]{Jedamzik:1996mr}%
  \BibitemOpen
  \bibfield  {author} {\bibinfo {author} {\bibfnamefont {K.}~\bibnamefont
  {Jedamzik}},\ }\href {\doibase 10.1103/PhysRevD.55.5871} {\bibfield
  {journal} {\bibinfo  {journal} {Phys.Rev.}\ }\textbf {\bibinfo {volume}
  {D55}},\ \bibinfo {pages} {5871} (\bibinfo {year} {1997})},\ \Eprint
  {http://arxiv.org/abs/astro-ph/9605152} {arXiv:astro-ph/9605152 [astro-ph]}
  \BibitemShut {NoStop}%
\bibitem [{\citenamefont {{Young}}\ and\ \citenamefont
  {{Byrnes}}(2013)}]{2013arXiv1307.4995Y}%
  \BibitemOpen
  \bibfield  {author} {\bibinfo {author} {\bibfnamefont {S.}~\bibnamefont
  {{Young}}}\ and\ \bibinfo {author} {\bibfnamefont {C.~T.}\ \bibnamefont
  {{Byrnes}}},\ }\href@noop {} {\bibfield  {journal} {\bibinfo  {journal}
  {ArXiv e-prints}\ } (\bibinfo {year} {2013})},\ \Eprint
  {http://arxiv.org/abs/1307.4995} {arXiv:1307.4995 [astro-ph.CO]} \BibitemShut
  {NoStop}%
\bibitem [{\citenamefont {Hawking}(1974)}]{Hawking:1974rv}%
  \BibitemOpen
  \bibfield  {author} {\bibinfo {author} {\bibfnamefont {S.}~\bibnamefont
  {Hawking}},\ }\href {\doibase 10.1038/248030a0} {\bibfield  {journal}
  {\bibinfo  {journal} {Nature}\ }\textbf {\bibinfo {volume} {248}},\ \bibinfo
  {pages} {30} (\bibinfo {year} {1974})}\BibitemShut {NoStop}%
\bibitem [{\citenamefont {Hawking}(1975)}]{Hawking:1974sw}%
  \BibitemOpen
  \bibfield  {author} {\bibinfo {author} {\bibfnamefont {S.}~\bibnamefont
  {Hawking}},\ }\href {\doibase 10.1007/BF02345020} {\bibfield  {journal}
  {\bibinfo  {journal} {Commun.Math.Phys.}\ }\textbf {\bibinfo {volume} {43}},\
  \bibinfo {pages} {199} (\bibinfo {year} {1975})}\BibitemShut {NoStop}%
\bibitem [{\citenamefont {Page}\ and\ \citenamefont
  {Hawking}(1976)}]{Page:1976wx}%
  \BibitemOpen
  \bibfield  {author} {\bibinfo {author} {\bibfnamefont {D.~N.}\ \bibnamefont
  {Page}}\ and\ \bibinfo {author} {\bibfnamefont {S.}~\bibnamefont {Hawking}},\
  }\href {\doibase 10.1086/154350} {\bibfield  {journal} {\bibinfo  {journal}
  {Astrophys.J.}\ }\textbf {\bibinfo {volume} {206}},\ \bibinfo {pages} {1}
  (\bibinfo {year} {1976})}\BibitemShut {NoStop}%
\bibitem [{\citenamefont {Quinn}\ \emph {et~al.}(2009)\citenamefont {Quinn},
  \citenamefont {Wilkinson}, \citenamefont {Irwin}, \citenamefont {Marshall},
  \citenamefont {Koch} \emph {et~al.}}]{Quinn:2009zg}%
  \BibitemOpen
  \bibfield  {author} {\bibinfo {author} {\bibfnamefont {D.}~\bibnamefont
  {Quinn}}, \bibinfo {author} {\bibfnamefont {M.}~\bibnamefont {Wilkinson}},
  \bibinfo {author} {\bibfnamefont {M.}~\bibnamefont {Irwin}}, \bibinfo
  {author} {\bibfnamefont {J.}~\bibnamefont {Marshall}}, \bibinfo {author}
  {\bibfnamefont {A.}~\bibnamefont {Koch}},  \emph {et~al.},\ }\href@noop {} {\
   (\bibinfo {year} {2009})},\ \Eprint {http://arxiv.org/abs/0903.1644}
  {arXiv:0903.1644 [astro-ph.GA]} \BibitemShut {NoStop}%
\bibitem [{\citenamefont {Carr}\ and\ \citenamefont
  {Sakellariadou}(1999)}]{0004-637X-516-1-195}%
  \BibitemOpen
  \bibfield  {author} {\bibinfo {author} {\bibfnamefont {B.~J.}\ \bibnamefont
  {Carr}}\ and\ \bibinfo {author} {\bibfnamefont {M.}~\bibnamefont
  {Sakellariadou}},\ }\href {http://stacks.iop.org/0004-637X/516/i=1/a=195}
  {\bibfield  {journal} {\bibinfo  {journal} {The Astrophysical Journal}\
  }\textbf {\bibinfo {volume} {516}},\ \bibinfo {pages} {195} (\bibinfo {year}
  {1999})}\BibitemShut {NoStop}%
\bibitem [{\citenamefont {Alcock}\ \emph {et~al.}(1998)\citenamefont {Alcock}
  \emph {et~al.}}]{MACHO}%
  \BibitemOpen
  \bibfield  {author} {\bibinfo {author} {\bibfnamefont {C.}~\bibnamefont
  {Alcock}} \emph {et~al.},\ }\href
  {http://stacks.iop.org/1538-4357/499/i=1/a=L9} {\bibfield  {journal}
  {\bibinfo  {journal} {The Astrophysical Journal Letters}\ }\textbf {\bibinfo
  {volume} {499}},\ \bibinfo {pages} {L9} (\bibinfo {year} {1998})}\BibitemShut
  {NoStop}%
\bibitem [{\citenamefont {{P. Tisserand}}\ \emph {et~al.}(2007)\citenamefont
  {{P. Tisserand}} \emph {et~al.}}]{EROS}%
  \BibitemOpen
  \bibfield  {author} {\bibinfo {author} {\bibnamefont {{P. Tisserand}}} \emph
  {et~al.},\ }\href {\doibase 10.1051/0004-6361:20066017} {\bibfield  {journal}
  {\bibinfo  {journal} {A\&A}\ }\textbf {\bibinfo {volume} {469}},\ \bibinfo
  {pages} {387} (\bibinfo {year} {2007})}\BibitemShut {NoStop}%
\bibitem [{\citenamefont {Marani}\ \emph {et~al.}(1999)\citenamefont {Marani},
  \citenamefont {Nemiroff}, \citenamefont {Norris}, \citenamefont {Hurley},\
  and\ \citenamefont {Bonnell}}]{GRB}%
  \BibitemOpen
  \bibfield  {author} {\bibinfo {author} {\bibfnamefont {G.~F.}\ \bibnamefont
  {Marani}}, \bibinfo {author} {\bibfnamefont {R.~J.}\ \bibnamefont
  {Nemiroff}}, \bibinfo {author} {\bibfnamefont {J.~P.}\ \bibnamefont
  {Norris}}, \bibinfo {author} {\bibfnamefont {K.}~\bibnamefont {Hurley}}, \
  and\ \bibinfo {author} {\bibfnamefont {J.~T.}\ \bibnamefont {Bonnell}},\
  }\href {http://stacks.iop.org/1538-4357/512/i=1/a=L13} {\bibfield  {journal}
  {\bibinfo  {journal} {The Astrophysical Journal Letters}\ }\textbf {\bibinfo
  {volume} {512}},\ \bibinfo {pages} {L13} (\bibinfo {year}
  {1999})}\BibitemShut {NoStop}%
\bibitem [{\citenamefont {{Dalcanton}}\ \emph {et~al.}(1994)\citenamefont
  {{Dalcanton}}, \citenamefont {{Canizares}}, \citenamefont {{Granados}},
  \citenamefont {{Steidel}},\ and\ \citenamefont {{Stocke}}}]{QSO}%
  \BibitemOpen
  \bibfield  {author} {\bibinfo {author} {\bibfnamefont {J.~J.}\ \bibnamefont
  {{Dalcanton}}}, \bibinfo {author} {\bibfnamefont {C.~R.}\ \bibnamefont
  {{Canizares}}}, \bibinfo {author} {\bibfnamefont {A.}~\bibnamefont
  {{Granados}}}, \bibinfo {author} {\bibfnamefont {C.~C.}\ \bibnamefont
  {{Steidel}}}, \ and\ \bibinfo {author} {\bibfnamefont {J.~T.}\ \bibnamefont
  {{Stocke}}},\ }\href {\doibase 10.1086/173914} {\bibfield  {journal}
  {\bibinfo  {journal} {\apj}\ }\textbf {\bibinfo {volume} {424}},\ \bibinfo
  {pages} {550} (\bibinfo {year} {1994})}\BibitemShut {NoStop}%
\bibitem [{\citenamefont {Ricotti}\ \emph {et~al.}(2007)\citenamefont
  {Ricotti}, \citenamefont {Ostriker},\ and\ \citenamefont
  {Mack}}]{Ricotti:2007au}%
  \BibitemOpen
  \bibfield  {author} {\bibinfo {author} {\bibfnamefont {M.}~\bibnamefont
  {Ricotti}}, \bibinfo {author} {\bibfnamefont {J.~P.}\ \bibnamefont
  {Ostriker}}, \ and\ \bibinfo {author} {\bibfnamefont {K.~J.}\ \bibnamefont
  {Mack}},\ }\href@noop {} {\  (\bibinfo {year} {2007})},\ \Eprint
  {http://arxiv.org/abs/0709.0524} {arXiv:0709.0524 [astro-ph]} \BibitemShut
  {NoStop}%
\bibitem [{\citenamefont {Mack}\ \emph {et~al.}(2007)\citenamefont {Mack},
  \citenamefont {Ostriker},\ and\ \citenamefont {Ricotti}}]{Mack:2006gz}%
  \BibitemOpen
  \bibfield  {author} {\bibinfo {author} {\bibfnamefont {K.~J.}\ \bibnamefont
  {Mack}}, \bibinfo {author} {\bibfnamefont {J.~P.}\ \bibnamefont {Ostriker}},
  \ and\ \bibinfo {author} {\bibfnamefont {M.}~\bibnamefont {Ricotti}},\ }\href
  {\doibase 10.1086/518998} {\bibfield  {journal} {\bibinfo  {journal}
  {Astrophys.J.}\ }\textbf {\bibinfo {volume} {665}},\ \bibinfo {pages} {1277}
  (\bibinfo {year} {2007})},\ \Eprint {http://arxiv.org/abs/astro-ph/0608642}
  {arXiv:astro-ph/0608642 [astro-ph]} \BibitemShut {NoStop}%
\bibitem [{\citenamefont {Kesden}\ and\ \citenamefont
  {Hanasoge}(2011)}]{Kesden:2011ij}%
  \BibitemOpen
  \bibfield  {author} {\bibinfo {author} {\bibfnamefont {M.}~\bibnamefont
  {Kesden}}\ and\ \bibinfo {author} {\bibfnamefont {S.}~\bibnamefont
  {Hanasoge}},\ }\href {\doibase 10.1103/PhysRevLett.107.111101} {\bibfield
  {journal} {\bibinfo  {journal} {Phys.Rev.Lett.}\ }\textbf {\bibinfo {volume}
  {107}},\ \bibinfo {pages} {111101} (\bibinfo {year} {2011})},\ \Eprint
  {http://arxiv.org/abs/1106.0011} {arXiv:1106.0011 [astro-ph.CO]} \BibitemShut
  {NoStop}%
\bibitem [{\citenamefont {Page}(1976)}]{PhysRevD.14.3260}%
  \BibitemOpen
  \bibfield  {author} {\bibinfo {author} {\bibfnamefont {D.~N.}\ \bibnamefont
  {Page}},\ }\href {\doibase 10.1103/PhysRevD.14.3260} {\bibfield  {journal}
  {\bibinfo  {journal} {Phys. Rev. D}\ }\textbf {\bibinfo {volume} {14}},\
  \bibinfo {pages} {3260} (\bibinfo {year} {1976})}\BibitemShut {NoStop}%
\bibitem [{\citenamefont {Frampton}\ \emph {et~al.}(2010)\citenamefont
  {Frampton}, \citenamefont {Kawasaki}, \citenamefont {Takahashi},\ and\
  \citenamefont {Yanagida}}]{Frampton:2010sw}%
  \BibitemOpen
  \bibfield  {author} {\bibinfo {author} {\bibfnamefont {P.~H.}\ \bibnamefont
  {Frampton}}, \bibinfo {author} {\bibfnamefont {M.}~\bibnamefont {Kawasaki}},
  \bibinfo {author} {\bibfnamefont {F.}~\bibnamefont {Takahashi}}, \ and\
  \bibinfo {author} {\bibfnamefont {T.~T.}\ \bibnamefont {Yanagida}},\ }\href
  {\doibase 10.1088/1475-7516/2010/04/023} {\bibfield  {journal} {\bibinfo
  {journal} {JCAP}\ }\textbf {\bibinfo {volume} {1004}},\ \bibinfo {pages}
  {023} (\bibinfo {year} {2010})},\ \Eprint {http://arxiv.org/abs/1001.2308}
  {arXiv:1001.2308 [hep-ph]} \BibitemShut {NoStop}%
\bibitem [{\citenamefont {Teukolsky}\ and\ \citenamefont
  {Press}(1974)}]{Teukolsky:1974yv}%
  \BibitemOpen
  \bibfield  {author} {\bibinfo {author} {\bibfnamefont {S.}~\bibnamefont
  {Teukolsky}}\ and\ \bibinfo {author} {\bibfnamefont {W.}~\bibnamefont
  {Press}},\ }\href {\doibase 10.1086/153180} {\bibfield  {journal} {\bibinfo
  {journal} {Astrophys.J.}\ }\textbf {\bibinfo {volume} {193}},\ \bibinfo
  {pages} {443} (\bibinfo {year} {1974})}\BibitemShut {NoStop}%
\bibitem [{\citenamefont {Press}\ and\ \citenamefont
  {Teukolsky}(1972)}]{Press:1972zz}%
  \BibitemOpen
  \bibfield  {author} {\bibinfo {author} {\bibfnamefont {W.~H.}\ \bibnamefont
  {Press}}\ and\ \bibinfo {author} {\bibfnamefont {S.~A.}\ \bibnamefont
  {Teukolsky}},\ }\href {\doibase 10.1038/238211a0} {\bibfield  {journal}
  {\bibinfo  {journal} {Nature}\ }\textbf {\bibinfo {volume} {238}},\ \bibinfo
  {pages} {211} (\bibinfo {year} {1972})}\BibitemShut {NoStop}%
\bibitem [{\citenamefont {Cardoso}(2013)}]{Cardoso:2013zfa}%
  \BibitemOpen
  \bibfield  {author} {\bibinfo {author} {\bibfnamefont {V.}~\bibnamefont
  {Cardoso}},\ }\href@noop {} {\  (\bibinfo {year} {2013})},\ \Eprint
  {http://arxiv.org/abs/1307.0038} {arXiv:1307.0038 [gr-qc]} \BibitemShut
  {NoStop}%
\bibitem [{\citenamefont {Damour}\ \emph {et~al.}(1976)\citenamefont {Damour},
  \citenamefont {Deruelle},\ and\ \citenamefont {Ruffini}}]{Damour:1976kh}%
  \BibitemOpen
  \bibfield  {author} {\bibinfo {author} {\bibfnamefont {T.}~\bibnamefont
  {Damour}}, \bibinfo {author} {\bibfnamefont {N.}~\bibnamefont {Deruelle}}, \
  and\ \bibinfo {author} {\bibfnamefont {R.}~\bibnamefont {Ruffini}},\
  }\href@noop {} {\bibfield  {journal} {\bibinfo  {journal} {Lett.Nuovo Cim.}\
  }\textbf {\bibinfo {volume} {15}},\ \bibinfo {pages} {257} (\bibinfo {year}
  {1976})}\BibitemShut {NoStop}%
\bibitem [{\citenamefont {Detweiler}(1980)}]{Detweiler:1980uk}%
  \BibitemOpen
  \bibfield  {author} {\bibinfo {author} {\bibfnamefont {S.~L.}\ \bibnamefont
  {Detweiler}},\ }\href {\doibase 10.1103/PhysRevD.22.2323} {\bibfield
  {journal} {\bibinfo  {journal} {Phys. Rev.}\ }\textbf {\bibinfo {volume}
  {D22}},\ \bibinfo {pages} {2323} (\bibinfo {year} {1980})}\BibitemShut
  {NoStop}%
\bibitem [{\citenamefont {Cardoso}\ \emph {et~al.}(2004)\citenamefont
  {Cardoso}, \citenamefont {Dias}, \citenamefont {Lemos},\ and\ \citenamefont
  {Yoshida}}]{Cardoso:2004nk}%
  \BibitemOpen
  \bibfield  {author} {\bibinfo {author} {\bibfnamefont {V.}~\bibnamefont
  {Cardoso}}, \bibinfo {author} {\bibfnamefont {O.~J.}\ \bibnamefont {Dias}},
  \bibinfo {author} {\bibfnamefont {J.~P.}\ \bibnamefont {Lemos}}, \ and\
  \bibinfo {author} {\bibfnamefont {S.}~\bibnamefont {Yoshida}},\ }\href
  {\doibase 10.1103/PhysRevD.70.044039, 10.1103/PhysRevD.70.049903} {\bibfield
  {journal} {\bibinfo  {journal} {Phys.Rev.}\ }\textbf {\bibinfo {volume}
  {D70}},\ \bibinfo {pages} {044039} (\bibinfo {year} {2004})},\ \Eprint
  {http://arxiv.org/abs/hep-th/0404096} {arXiv:hep-th/0404096 [hep-th]}
  \BibitemShut {NoStop}%
\bibitem [{\citenamefont {Sitenko}(1976)}]{Sitenko:1967}%
  \BibitemOpen
  \bibfield  {author} {\bibinfo {author} {\bibfnamefont {A.~G.}\ \bibnamefont
  {Sitenko}},\ }\href@noop {} {\emph {\bibinfo {title} {{Electromagnetic
  Fluctuations in Plasma}}}}\ (\bibinfo  {publisher} {Academic, New York},\
  \bibinfo {year} {1976})\BibitemShut {NoStop}%
\bibitem [{\citenamefont {Kulsrud}\ and\ \citenamefont
  {Loeb}(1992)}]{Kulsrud:1991jt}%
  \BibitemOpen
  \bibfield  {author} {\bibinfo {author} {\bibfnamefont {R.}~\bibnamefont
  {Kulsrud}}\ and\ \bibinfo {author} {\bibfnamefont {A.}~\bibnamefont {Loeb}},\
  }\href {\doibase 10.1103/PhysRevD.45.525} {\bibfield  {journal} {\bibinfo
  {journal} {Phys.Rev.}\ }\textbf {\bibinfo {volume} {D45}},\ \bibinfo {pages}
  {525} (\bibinfo {year} {1992})}\BibitemShut {NoStop}%
\bibitem [{\citenamefont {Arvanitaki}\ and\ \citenamefont
  {Dubovsky}(2011)}]{Arvanitaki:2010sy}%
  \BibitemOpen
  \bibfield  {author} {\bibinfo {author} {\bibfnamefont {A.}~\bibnamefont
  {Arvanitaki}}\ and\ \bibinfo {author} {\bibfnamefont {S.}~\bibnamefont
  {Dubovsky}},\ }\href {\doibase 10.1103/PhysRevD.83.044026} {\bibfield
  {journal} {\bibinfo  {journal} {Phys.Rev.}\ }\textbf {\bibinfo {volume}
  {D83}},\ \bibinfo {pages} {044026} (\bibinfo {year} {2011})},\ \Eprint
  {http://arxiv.org/abs/1004.3558} {arXiv:1004.3558 [hep-th]} \BibitemShut
  {NoStop}%
\bibitem [{\citenamefont {Sunyaev}\ and\ \citenamefont
  {Khatri}(2013)}]{Sunyaev:2013aoa}%
  \BibitemOpen
  \bibfield  {author} {\bibinfo {author} {\bibfnamefont {R.~A.}\ \bibnamefont
  {Sunyaev}}\ and\ \bibinfo {author} {\bibfnamefont {R.}~\bibnamefont
  {Khatri}},\ }\href {\doibase 10.1142/S0218271813300140} {\bibfield  {journal}
  {\bibinfo  {journal} {Int.J.Mod.Phys.}\ }\textbf {\bibinfo {volume} {D22}},\
  \bibinfo {pages} {1330014} (\bibinfo {year} {2013})},\ \Eprint
  {http://arxiv.org/abs/1302.6553} {arXiv:1302.6553 [astro-ph.CO]} \BibitemShut
  {NoStop}%
\bibitem [{\citenamefont {Fixsen}\ \emph {et~al.}(1996)\citenamefont {Fixsen},
  \citenamefont {Cheng}, \citenamefont {Gales}, \citenamefont {Mather},
  \citenamefont {Shafer} \emph {et~al.}}]{Fixsen:1996nj}%
  \BibitemOpen
  \bibfield  {author} {\bibinfo {author} {\bibfnamefont {D.}~\bibnamefont
  {Fixsen}}, \bibinfo {author} {\bibfnamefont {E.}~\bibnamefont {Cheng}},
  \bibinfo {author} {\bibfnamefont {J.}~\bibnamefont {Gales}}, \bibinfo
  {author} {\bibfnamefont {J.~C.}\ \bibnamefont {Mather}}, \bibinfo {author}
  {\bibfnamefont {R.}~\bibnamefont {Shafer}},  \emph {et~al.},\ }\href
  {\doibase 10.1086/178173} {\bibfield  {journal} {\bibinfo  {journal}
  {Astrophys.J.}\ }\textbf {\bibinfo {volume} {473}},\ \bibinfo {pages} {576}
  (\bibinfo {year} {1996})},\ \Eprint {http://arxiv.org/abs/astro-ph/9605054}
  {arXiv:astro-ph/9605054 [astro-ph]} \BibitemShut {NoStop}%
\bibitem [{\citenamefont {Kogut}\ \emph {et~al.}(2011)\citenamefont {Kogut},
  \citenamefont {Fixsen}, \citenamefont {Chuss}, \citenamefont {Dotson},
  \citenamefont {Dwek} \emph {et~al.}}]{Kogut:2011xw}%
  \BibitemOpen
  \bibfield  {author} {\bibinfo {author} {\bibfnamefont {A.}~\bibnamefont
  {Kogut}}, \bibinfo {author} {\bibfnamefont {D.}~\bibnamefont {Fixsen}},
  \bibinfo {author} {\bibfnamefont {D.}~\bibnamefont {Chuss}}, \bibinfo
  {author} {\bibfnamefont {J.}~\bibnamefont {Dotson}}, \bibinfo {author}
  {\bibfnamefont {E.}~\bibnamefont {Dwek}},  \emph {et~al.},\ }\href {\doibase
  10.1088/1475-7516/2011/07/025} {\bibfield  {journal} {\bibinfo  {journal}
  {JCAP}\ }\textbf {\bibinfo {volume} {1107}},\ \bibinfo {pages} {025}
  (\bibinfo {year} {2011})},\ \Eprint {http://arxiv.org/abs/1105.2044}
  {arXiv:1105.2044 [astro-ph.CO]} \BibitemShut {NoStop}%
\bibitem [{\citenamefont {Pani}\ \emph
  {et~al.}(2012{\natexlab{a}})\citenamefont {Pani}, \citenamefont {Cardoso},
  \citenamefont {Gualtieri}, \citenamefont {Berti},\ and\ \citenamefont
  {Ishibashi}}]{Pani:2012bp}%
  \BibitemOpen
  \bibfield  {author} {\bibinfo {author} {\bibfnamefont {P.}~\bibnamefont
  {Pani}}, \bibinfo {author} {\bibfnamefont {V.}~\bibnamefont {Cardoso}},
  \bibinfo {author} {\bibfnamefont {L.}~\bibnamefont {Gualtieri}}, \bibinfo
  {author} {\bibfnamefont {E.}~\bibnamefont {Berti}}, \ and\ \bibinfo {author}
  {\bibfnamefont {A.}~\bibnamefont {Ishibashi}},\ }\href {\doibase
  10.1103/PhysRevD.86.104017} {\bibfield  {journal} {\bibinfo  {journal}
  {Phys.Rev.}\ }\textbf {\bibinfo {volume} {D86}},\ \bibinfo {pages} {104017}
  (\bibinfo {year} {2012}{\natexlab{a}})},\ \Eprint
  {http://arxiv.org/abs/1209.0773} {arXiv:1209.0773 [gr-qc]} \BibitemShut
  {NoStop}%
\bibitem [{\citenamefont {Pani}\ \emph
  {et~al.}(2012{\natexlab{b}})\citenamefont {Pani}, \citenamefont {Cardoso},
  \citenamefont {Gualtieri}, \citenamefont {Berti},\ and\ \citenamefont
  {Ishibashi}}]{Pani:2012vp}%
  \BibitemOpen
  \bibfield  {author} {\bibinfo {author} {\bibfnamefont {P.}~\bibnamefont
  {Pani}}, \bibinfo {author} {\bibfnamefont {V.}~\bibnamefont {Cardoso}},
  \bibinfo {author} {\bibfnamefont {L.}~\bibnamefont {Gualtieri}}, \bibinfo
  {author} {\bibfnamefont {E.}~\bibnamefont {Berti}}, \ and\ \bibinfo {author}
  {\bibfnamefont {A.}~\bibnamefont {Ishibashi}},\ }\href {\doibase
  10.1103/PhysRevLett.109.131102} {\bibfield  {journal} {\bibinfo  {journal}
  {Phys.Rev.Lett.}\ }\textbf {\bibinfo {volume} {109}},\ \bibinfo {pages}
  {131102} (\bibinfo {year} {2012}{\natexlab{b}})},\ \Eprint
  {http://arxiv.org/abs/1209.0465} {arXiv:1209.0465 [gr-qc]} \BibitemShut
  {NoStop}%
\bibitem [{\citenamefont {Witek}\ \emph {et~al.}(2013)\citenamefont {Witek},
  \citenamefont {Cardoso}, \citenamefont {Ishibashi},\ and\ \citenamefont
  {Sperhake}}]{Witek:2012tr}%
  \BibitemOpen
  \bibfield  {author} {\bibinfo {author} {\bibfnamefont {H.}~\bibnamefont
  {Witek}}, \bibinfo {author} {\bibfnamefont {V.}~\bibnamefont {Cardoso}},
  \bibinfo {author} {\bibfnamefont {A.}~\bibnamefont {Ishibashi}}, \ and\
  \bibinfo {author} {\bibfnamefont {U.}~\bibnamefont {Sperhake}},\ }\href
  {\doibase 10.1103/PhysRevD.87.043513} {\bibfield  {journal} {\bibinfo
  {journal} {Phys.Rev.}\ }\textbf {\bibinfo {volume} {D87}},\ \bibinfo {pages}
  {043513} (\bibinfo {year} {2013})},\ \Eprint {http://arxiv.org/abs/1212.0551}
  {arXiv:1212.0551 [gr-qc]} \BibitemShut {NoStop}%
\bibitem [{\citenamefont {Dolan}(2007)}]{Dolan:2007mj}%
  \BibitemOpen
  \bibfield  {author} {\bibinfo {author} {\bibfnamefont {S.~R.}\ \bibnamefont
  {Dolan}},\ }\href {\doibase 10.1103/PhysRevD.76.084001} {\bibfield  {journal}
  {\bibinfo  {journal} {Phys.Rev.}\ }\textbf {\bibinfo {volume} {D76}},\
  \bibinfo {pages} {084001} (\bibinfo {year} {2007})},\ \Eprint
  {http://arxiv.org/abs/0705.2880} {arXiv:0705.2880 [gr-qc]} \BibitemShut
  {NoStop}%
\bibitem [{\citenamefont {Wald}(1984)}]{Wald:1984rg}%
  \BibitemOpen
  \bibfield  {author} {\bibinfo {author} {\bibfnamefont {R.~M.}\ \bibnamefont
  {Wald}},\ }\href@noop {} {\emph {\bibinfo {title} {{General Relativity}}}}\
  (\bibinfo  {publisher} {University Of Chicago Press},\ \bibinfo {year}
  {1984})\BibitemShut {NoStop}%
\bibitem [{\citenamefont {Ade}\ \emph {et~al.}(2013)\citenamefont {Ade} \emph
  {et~al.}}]{Ade:2013zuv}%
  \BibitemOpen
  \bibfield  {author} {\bibinfo {author} {\bibfnamefont {P.}~\bibnamefont
  {Ade}} \emph {et~al.} (\bibinfo {collaboration} {Planck Collaboration}),\
  }\href@noop {} {\  (\bibinfo {year} {2013})},\ \Eprint
  {http://arxiv.org/abs/1303.5076} {arXiv:1303.5076 [astro-ph.CO]} \BibitemShut
  {NoStop}%
\bibitem [{\citenamefont {Kuzelev}\ and\ \citenamefont
  {Rukhadze}(2009)}]{Kuzelev:2009}%
  \BibitemOpen
  \bibfield  {author} {\bibinfo {author} {\bibfnamefont {M.~V.}\ \bibnamefont
  {Kuzelev}}\ and\ \bibinfo {author} {\bibfnamefont {A.~A.}\ \bibnamefont
  {Rukhadze}},\ }\href@noop {} {\emph {\bibinfo {title} {{Methods of Wave
  Theory in Dispersive Media}}}}\ (\bibinfo  {publisher} {World Scientific,
  Singapore},\ \bibinfo {year} {2009})\BibitemShut {NoStop}%
\bibitem [{\citenamefont {{Burigana}}\ \emph {et~al.}(1991)\citenamefont
  {{Burigana}}, \citenamefont {{Danese}},\ and\ \citenamefont {{de
  Zotti}}}]{1991ApJ...379....1B}%
  \BibitemOpen
  \bibfield  {author} {\bibinfo {author} {\bibfnamefont {C.}~\bibnamefont
  {{Burigana}}}, \bibinfo {author} {\bibfnamefont {L.}~\bibnamefont
  {{Danese}}}, \ and\ \bibinfo {author} {\bibfnamefont {G.}~\bibnamefont {{de
  Zotti}}},\ }\href {\doibase 10.1086/170479} {\bibfield  {journal} {\bibinfo
  {journal} {\apj}\ }\textbf {\bibinfo {volume} {379}},\ \bibinfo {pages} {1}
  (\bibinfo {year} {1991})}\BibitemShut {NoStop}%
\bibitem [{\citenamefont {Khatri}\ and\ \citenamefont
  {Sunyaev}(2013)}]{Khatri:2013dha}%
  \BibitemOpen
  \bibfield  {author} {\bibinfo {author} {\bibfnamefont {R.}~\bibnamefont
  {Khatri}}\ and\ \bibinfo {author} {\bibfnamefont {R.~A.}\ \bibnamefont
  {Sunyaev}},\ }\href {\doibase 10.1088/1475-7516/2013/06/026} {\bibfield
  {journal} {\bibinfo  {journal} {JCAP}\ }\textbf {\bibinfo {volume} {1306}},\
  \bibinfo {pages} {026} (\bibinfo {year} {2013})},\ \Eprint
  {http://arxiv.org/abs/1303.7212} {arXiv:1303.7212 [astro-ph.CO]} \BibitemShut
  {NoStop}%
\bibitem [{\citenamefont {Chluba}\ and\ \citenamefont
  {Jeong}(2013)}]{Chluba:2013dza}%
  \BibitemOpen
  \bibfield  {author} {\bibinfo {author} {\bibfnamefont {J.}~\bibnamefont
  {Chluba}}\ and\ \bibinfo {author} {\bibfnamefont {D.}~\bibnamefont {Jeong}},\
  }\href@noop {} {\  (\bibinfo {year} {2013})},\ \Eprint
  {http://arxiv.org/abs/1306.5751} {arXiv:1306.5751 [astro-ph.CO]} \BibitemShut
  {NoStop}%
\bibitem [{\citenamefont {{Griest}}\ \emph {et~al.}(2011)\citenamefont
  {{Griest}}, \citenamefont {{Lehner}}, \citenamefont {{Cieplak}},\ and\
  \citenamefont {{Jain}}}]{2011PhRvL.107w1101G}%
  \BibitemOpen
  \bibfield  {author} {\bibinfo {author} {\bibfnamefont {K.}~\bibnamefont
  {{Griest}}}, \bibinfo {author} {\bibfnamefont {M.~J.}\ \bibnamefont
  {{Lehner}}}, \bibinfo {author} {\bibfnamefont {A.~M.}\ \bibnamefont
  {{Cieplak}}}, \ and\ \bibinfo {author} {\bibfnamefont {B.}~\bibnamefont
  {{Jain}}},\ }\href {\doibase 10.1103/PhysRevLett.107.231101} {\bibfield
  {journal} {\bibinfo  {journal} {Physical Review Letters}\ }\textbf {\bibinfo
  {volume} {107}},\ \bibinfo {eid} {231101} (\bibinfo {year} {2011})},\ \Eprint
  {http://arxiv.org/abs/1109.4975} {arXiv:1109.4975 [astro-ph.CO]} \BibitemShut
  {NoStop}%
\bibitem [{\citenamefont {{Cieplak}}\ and\ \citenamefont
  {{Griest}}(2013)}]{2013ApJ...767..145C}%
  \BibitemOpen
  \bibfield  {author} {\bibinfo {author} {\bibfnamefont {A.~M.}\ \bibnamefont
  {{Cieplak}}}\ and\ \bibinfo {author} {\bibfnamefont {K.}~\bibnamefont
  {{Griest}}},\ }\href {\doibase 10.1088/0004-637X/767/2/145} {\bibfield
  {journal} {\bibinfo  {journal} {\apj}\ }\textbf {\bibinfo {volume} {767}},\
  \bibinfo {eid} {145} (\bibinfo {year} {2013})},\ \Eprint
  {http://arxiv.org/abs/1210.7729} {arXiv:1210.7729 [astro-ph.CO]} \BibitemShut
  {NoStop}%
\bibitem [{\citenamefont {Kashiyama}\ and\ \citenamefont
  {Seto}(2012)}]{Kashiyama:2012qz}%
  \BibitemOpen
  \bibfield  {author} {\bibinfo {author} {\bibfnamefont {K.}~\bibnamefont
  {Kashiyama}}\ and\ \bibinfo {author} {\bibfnamefont {N.}~\bibnamefont
  {Seto}},\ }\href@noop {} {\  (\bibinfo {year} {2012})},\ \Eprint
  {http://arxiv.org/abs/1208.4101} {arXiv:1208.4101 [astro-ph.CO]} \BibitemShut
  {NoStop}%
\bibitem [{\citenamefont {Capela}\ \emph {et~al.}(2013)\citenamefont {Capela},
  \citenamefont {Pshirkov},\ and\ \citenamefont {Tinyakov}}]{Capela:2013yf}%
  \BibitemOpen
  \bibfield  {author} {\bibinfo {author} {\bibfnamefont {F.}~\bibnamefont
  {Capela}}, \bibinfo {author} {\bibfnamefont {M.}~\bibnamefont {Pshirkov}}, \
  and\ \bibinfo {author} {\bibfnamefont {P.}~\bibnamefont {Tinyakov}},\ }\href
  {\doibase 10.1103/PhysRevD.87.123524} {\bibfield  {journal} {\bibinfo
  {journal} {Phys. Rev. D 87,}\ }\textbf {\bibinfo {volume} {123524}} (\bibinfo
  {year} {2013}),\ 10.1103/PhysRevD.87.123524},\ \Eprint
  {http://arxiv.org/abs/1301.4984} {arXiv:1301.4984 [astro-ph.CO]} \BibitemShut
  {NoStop}%
\bibitem [{\citenamefont {Pani}(2013)}]{Pani:2013pma}%
  \BibitemOpen
  \bibfield  {author} {\bibinfo {author} {\bibfnamefont {P.}~\bibnamefont
  {Pani}},\ }\href@noop {} {\  (\bibinfo {year} {2013})},\ \Eprint
  {http://arxiv.org/abs/1305.6759} {arXiv:1305.6759 [gr-qc]} \BibitemShut
  {NoStop}%
\bibitem [{\citenamefont {Dolan}(2013)}]{Dolan:2012yt}%
  \BibitemOpen
  \bibfield  {author} {\bibinfo {author} {\bibfnamefont {S.~R.}\ \bibnamefont
  {Dolan}},\ }\href {\doibase 10.1103/PhysRevD.87.124026} {\bibfield  {journal}
  {\bibinfo  {journal} {Phys. Rev. D 87,}\ }\textbf {\bibinfo {volume}
  {124026}} (\bibinfo {year} {2013}),\ 10.1103/PhysRevD.87.124026},\ \Eprint
  {http://arxiv.org/abs/1212.1477} {arXiv:1212.1477 [gr-qc]} \BibitemShut
  {NoStop}%
\end{thebibliography}%
\end{document}